\documentclass[epj]{svjour}
\usepackage{graphics}
\usepackage{graphicx}
\usepackage{dcolumn}
\usepackage{bm}
\usepackage{mathrsfs}
\usepackage{amsmath}
\usepackage{amssymb}
\usepackage{float}
\usepackage{dcolumn}
\usepackage{multirow}
\usepackage{setspace}
\begin{document}
\title{Possible open-charmed pentaquark molecule $\Omega_c(3188)$ --- the $D \Xi$ bound state --- in the Bethe-Salpeter formalism}
\author{Chao Wang\inst{1} \and Liang-Liang Liu \inst{2}\and Xian-Wei Kang \inst{3,4}\and Xin-Heng Guo \inst{3}\thanks{\emph{Corresponding author, e-mail:} xhguo@bnu.edu.cn} \and Rui-Wu Wang \inst{1}
\thanks{\emph{Corresponding author, e-mail:} wangrw@nwpu.edu.cn}%
}                     
%
%
\institute{Center for Ecological and Environmental Sciences, Key Laboratory for Space Bioscience $\&$ Biotechnology, Northwestern Polytechnical University, Xi'an 710072, China \and College of Physics and Information Engineering, Shanxi Normal University, Linfen 041004, China \and College of Nuclear Science and Technology,
Beijing Normal University, Beijing 100875, China\and  Institute of Physics, Academia Sinica, Taipei, Taiwan 115}
\date{Received: date / Revised version: date}
%
\abstract{
We study the $S$-wave $D\Xi$ bound state in the Bethe-Salpeter formalism in the ladder and instantaneous approximations. With the kernel generated by the hadronic effective Lagrangian, two open-charmed bound states, which quantum numbers are $I=0$,~$J^P=(\frac{1}{2})^-$ and $I=1$,~$J^P=(\frac{1}{2})^-$, respectively, are predicted as new candidates of hadronic pentaquark molecules in our formalism. If existing, they could contribute to the broad 3188\,MeV structure near the five new narrow $\Omega_c$ states observed recently by the LHCb Collaboration.
\PACS{{14.20.Pt}{Exotic baryons} \and
      {12.40.Yx}{Hadron mass models and calculations} \and
      {11.10.St}{Bound and unstable states; Bethe-Salpeter equations}
     } 
}
\titlerunning{Possible open-charmed pentaquark molecule $D\Xi$ in the Bethe-Salpeter formalism}
\authorrunning{Chao Wang et al.}
\maketitle

\section{Introduction}
\label{intro}
{\baselineskip=0.985\baselineskip
In the last decade, many important experimental progresses were made in the study of charm hadrons. Several charm baryons and their excited states have been reported and this stimulates great interest in understanding the structures of charm baryons. In these experimental observations, there are some non-conventional states, which are more complicated than the hadronic states in the classical quark model. These states can be interpreted as five-quark or meson-baryon bound states. In 2015, the LHCb Collaboration discovered two hidden-charmed pentaquark-like structures $P_c(4380)^+$ and $P_c(4450)^+$ \cite{Aaij:2015tga,Aaij:2016phn,Aaij:2016ymb}, which are considered as $\Sigma_c\bar D^*$, $\Sigma_c^*\bar D$, or $\Sigma_c^*\bar D^*$ pentaquark molecule \cite{Chen:2015loa,Chen:2015moa,Roca:2015dva,He:2015cea}. Besides the hidden-charmed pantaquarks, only a few open-charmed baryons are treated as candidates for pentaquark molecules which have not yet been confirmed by experiments. For example, $\Sigma_c(2800)$ and $\Lambda_c(2940)^+$ have been suggested to be $S$-wave $DN$ and $D^*N$ molecular states, respectively \cite{Dong:2010xv,He:2010zq,Zhang:2012jk,Aaij:2017vbw}.

Very recently, the LHCb Collaboration declared that they observed five new narrow $\Omega_c$ states, which are $\Omega_c(3000)^0$, $\Omega_c(3050)^0$, $\Omega_c(3066)^0$, $\Omega_c(3090)^0$, and $\Omega_c(3119)^0$ \cite{Aaij:2017nav}. Actually, there are six structures existing in the spectrum of invariant mass of $\Xi^+_cK^-$. The data also indicate the presence of a broad structure around 3188\,MeV that is fitted as single resonance [shortly denoted by $\Omega_c(3188)$ in our work]. In the experiment, this resonance is described as the sum of four incoherent Breit-Wigner functions to be left into the systematic uncertainties of the others five resonances \cite{Aaij:2017nav}. However, one notes that this structure is located at the $D\Xi$ threshold (3179-3191\,MeV) and therefore can also be interpreted as the $D\Xi$ bound state naturally. At the same time, since the width of corresponding peak is quite broad \cite{Aaij:2017nav}, if this structure is a real signal of resonance, there is reason to believe that it would be a new pentaquark molecule with a single charm quark. At present, there is no enough information concerning this structure, such as spin-parity, it is interesting and significative to make efforts on the theoretical side to confirm its existence and reveal its properties. In this work, we will focus on this structure and study the possible $S$-wave $D\Xi$ bound state.

The Bethe-Salpeter (BS) equation is a formally exact equation to describe the relativistic bound state. This technique was developed by Feynman, Bethe, and  Salpeter et al. \cite{Feynman:1949hz,Salpeter:1951sz,DL}. It has been applied to theoretical studies concerning heavy baryons and molecular bound states \cite{Guo:1996jj,Guo:2007mm,Xie:2010zza,Feng:2011zzb,Wang:2017rjs,Ke:2016oez}. In previous studies, the possible bound states of $K\bar K$, $DK$, $B\bar K$, and $K^-p$ have been investigated in the BS formalism in the ladder and instantaneous approximations \cite{Guo:2007mm,Xie:2010zza,Feng:2011zzb,Wang:2017rjs}. We will try to study $S$-wave $D\Xi$ molecular bound state with the kernel introduced by the vector meson exchange interactions in this framework. We will investigate whether this state exists or not and study its decay. We will also discuss the possibility of the $\Omega_c(3188)$ structure to be the $D\Xi$ bound state.

The remainder of this paper is organized as follows. In Sect.~II, we give the generalized formalism of the BS equation for the fermion-scalar system. In Sect.~III, we derive the BS equation for the $D\Xi$ system in detail and present the normalization condition of the corresponding BS amplitude. In Sect.~III, the decay of the $D\Xi$ bound state to $\Xi_c^+K^-$ is discussed. The numerical results are presented in Sect.~IV. In the last section, we give a summary and some discussions.

\section{The Bethe-Salpeter equation for the fermion-scalar system }
In this section, we will present the BS equation for the fermion-scalar system. We assume that the bound state exists in a fermion-scalar system and its mass is $M$. The BS wave function can be defined as \cite{DL,Guo:1996jj,Guo:2007mm,Xie:2010zza,Feng:2011zzb,Wang:2017rjs}
\begin{eqnarray}
\chi(x_1,x_2,P)&=&\langle 0|T\psi(x_1) \phi(x_2)|P\rangle,
\end{eqnarray}
with $\psi(x_{1})$ and $\phi(x_{2})$ being field operators of the fermion and scalar particles , respectively, and $P$ being the momentum of the system. In momentum space, the BS wave function, $\chi_{P}(p)$, is related to $\chi(x_1,x_2,P)$ through the following equation \cite{DL}:
\begin{eqnarray}
\chi(x_{1},x_{2},P)=\mathrm e^{\mathrm iPX}\int\frac{\mathrm{d}^{4}p}{(2\pi)^4}\chi_{P}(p)\mathrm e^{\mathrm ipx},
\end{eqnarray}
where $p$ and $x(=x_{1}-x_{2})$ are the relative momentum and the relative coordinate of two constituents, respectively, and $X$ is the center of mass coordinate which is defined as $X=\lambda_{1}x_{1}+\lambda_{2}x_{2}$, where $\lambda_{1}=\frac{m_{1}}{m_{1}+m_{2}}$, $\lambda_{2}=\frac{m_{2}}{m_{1}+m_{2}}$, with $m_{1}$ and $m_{2}$ being the masses of the fermion and the scalar constituent, respectively. The momentum of the fermion is $p_{1}=\lambda_{1}P+p$ and that of the scalar particle is $p_{2}=\lambda_{2}P-p$. The derivation of the BS formalism for the two fermion system can be found in the textbook \cite{DL}. In the same way, one can prove that the form of the BS equation is still valid for the fermion and scalar object system. The BS wave function in our case satisfies the follow homogeneous integral equation \cite{DL,Guo:1996jj,Guo:2007mm,Xie:2010zza,Feng:2011zzb,Wang:2017rjs}:
\begin{eqnarray}\label{BSe1.1}
\chi_{P}(p)&=&s_{F}(\lambda_{1}P+p)\int\frac{\mathrm d^{4}q}{(2\pi)^4}K(P,\,p,\,q) \chi_{P}(q)\nonumber\\[4pt]
&&\qquad \qquad \times s_{S}(\lambda_{2}P-p),
\end{eqnarray}
where $s_F$ and $s_S$ are propagators of the fermion and the scalar particle, respectively, and $K(P,p,q)$ is the interaction kernel which can be described by the sum of all the irreducible graphs which cannot be split into two pieces by cutting two particle lines as defined in Ref.~\cite{DL}. For convenience, we also define the relative longitudinal momentum $p_l(=v\cdot p)$ and the relative transverse momentum $p_t[=p-(v\cdot p)v]$ with $v(=P/M)$ being the four velocity of the bound state.

\section{The Bethe-Salpeter formalism for the $D\Xi$ bound state}

\subsection{Isospin structure of the possible $D \Xi$ bound state}
The isospin field doublets, $D=(-D^+,\,D^0)^T$ and $\Xi=(\Xi^0,\,\Xi^-)^T$, have the following expansions in momentum space:
\begin{eqnarray}
D_1(x)&=&\int\frac{\mathrm d^3 p}{(2\pi)^3\sqrt{2E_{\bf p}}}\left(a_{D^+}\mathrm e^{-\mathrm i px}+a_{D^-}^\dag\mathrm e^{\mathrm i px}\right),\nonumber\\
D_2(x)&=&\int\frac{\mathrm d^3 p}{(2\pi)^3\sqrt{2E_{\bf p}}}\left(a_{D^0 }\mathrm e^{-\mathrm i px}+a_{\bar{D}^0}^\dag \mathrm e^{\mathrm i px}\right),\nonumber\\
\Xi_1(x)&=&\int\frac{\mathrm d^3 p}{(2\pi)^3\sqrt{2E_{\bf p}}}\big[a_{\Xi^0} \mathrm e^{-\mathrm i px}u_\Xi(p)+a^\dag_{\bar \Xi^0} \mathrm e^{\mathrm i px}v_\Xi(p)\big],\nonumber\\
\Xi_2(x)&=&\int\frac{\mathrm d^3 p}{(2\pi)^3\sqrt{2E_{\bf p}}}\big[a_{\Xi^-} \mathrm e^{-\mathrm i px}u_\Xi(p)+a_{\Xi^+}^{\dag} \mathrm e^{\mathrm i px}v_\Xi(p)\big],\nonumber\\
\end{eqnarray}
with $E_{\bf p}=\sqrt{|{\bf p}|^2+m^2}$.

The isoscalar bound state can be written as
\begin{eqnarray}
|P\rangle_{(0,\,0)}=\frac{1}{\sqrt 2}|D^0\Xi^0+D^+\Xi^-\rangle,
\end{eqnarray}
while the isovector one is
\begin{eqnarray}
&&|P\rangle_{(1,\,0)}=\frac{1}{\sqrt 2}|D^0\Xi^0-D^+\Xi^-\rangle,\nonumber\\
&&|P\rangle_{(1,\,1)}=-|D^+\Xi^0\rangle,\nonumber\\
&&|P\rangle_{(1,\,-1)}=|D^0\Xi^-\rangle,
\end{eqnarray}
where the subscripts $(I,\,I_3)$ denote the isospin and the third component of the isospin. Since the interaction between the two constituents is dominated by strong interaction, the BS wave function depends only on the isospin $I$. Therefore, the BS wave function of the $D\Xi$ system can be defined as
\begin{eqnarray}
\langle 0| \text{T}\{D_i(x_1)\Xi_j(x_2)\} |P \rangle^{I}=C^{ij}_{(I,\,I_3)}\chi^I_P(x_1,\,x_2),
\end{eqnarray}
where the isospin coefficients $C^{ij}_{(I,\,I_3)}$ are
\begin{align}
C^{12}_{(0,\,0)}=C^{21}_{(0,\,0)}=\frac{1}{\sqrt 2},\qquad  \text{else}=0,
\end{align}
for the isoscalar state and
\begin{alignat}{3}
C^{12}_{(1,\,0)}&=-\frac{1}{\sqrt{2}}, \qquad &  &C^{21}_{(1,\,0)}=\frac{1}{\sqrt{2}},\qquad&\nonumber\\
C^{12}_{(1,\,1)}&=-1,                         &  &C^{21}_{(1,\,-1)}=1,& \text{else}&=0,
\end{alignat}
for the isovector state.

Considering the isospin structure, the BS equation for the $D\Xi$ system can be written as
\begin{eqnarray}\label{BSES}
C^{ij}_{(I)}\chi^I_{P}(p)&=&s_{\Xi}(\lambda_{1}P+p) \int\frac{\mathrm d^{4}q}{(2\pi)^4}K^{ij,\,lk}(P,\,p,\,q)\nonumber\\[4pt]
&&\qquad \qquad \times C^{lk}_{(I)} \chi^I_{P}(q)s_{D}(\lambda_{2}P-p),
\end{eqnarray}
where $i(j)$ and $l(k)$ refer to the components of the $D(\Xi)$ field doublets, we neglect the effect of isospin violation and take $m_\Xi=\frac{1}{2}(m_{\Xi^0}+m_{\Xi^-})$ and $m_D=\frac{1}{2}(m_{D^+}+m_{D^0})$. Explicitly, we give the isoscalar case as an example:
\begin{eqnarray}
\chi^{I=0}_{P}(p)&=&s_{\Xi}(\lambda_{1}P+p)  \nonumber\\[4pt]
&&\times \int\frac{\mathrm d^{4}q}{(2\pi)^4} \left [K^{12,\,12}(P,\,p,\,q)+ K^{12,\,21}(P,\,p,\,q)\right] \nonumber\\[4pt]
&&\qquad \qquad \qquad \times \chi^{I=0}_{P}(q)s_{D}(\lambda_{2}P-p),
\end{eqnarray}
in which we can see that $K(P,\,p,\,q)=K^{12,\,12}(P,\,p,\,q)+ K^{12,\,21}(P,\,p,\,q)$.
\subsection{The Bethe-Salpeter equation for the $D\Xi$ bound state}
In general, considering $v\!\!\!/u(v, s) = u(v, s)$ , $\chi_P(p)$ can be written as \cite{Zhang:2013gqa,Liu:2015qfa,Liu:2016wzh}
\begin{eqnarray}
\chi_{P}(p)=(g_1+g_2\gamma_5+g_3\gamma_5p\!\!\!/_t+g_4p\!\!\!/_t)u(v,s),
\end{eqnarray}
where $u(v,s)$ is the spinor of the bound state with helicity $s$ and $g_i$($i=1,2\cdots4$) are Lorentz-scalar functions. According to our previous works, the momentum transfer between the two constituents in a hadronic bound state is in the order of 0.1\,GeV \cite{Guo:2007mm,Xie:2010zza,Feng:2011zzb,Wang:2017rjs} and this quantity is quite smaller than the mass of $D$. Thus, we can apply the heavy quark symmetry and have $v\!\!\!/\chi_{P}(p) = \chi_{P}(p)$. With the constraints imposed by parity and Lorentz transformations, it is easy to prove that $\chi_{P}(p)$ can be simplified as \cite{Zhang:2013gqa,Liu:2015qfa}
\begin{eqnarray}\label{BSA}
\chi_{P}(p)=f(p)u(v,s),
\end{eqnarray}
in which $f(p)$ is a Lorentz-scalar function of $p$.

\begin{figure}[bt]
\centering
\includegraphics[bb=196 583 379 712, width=0.35\textwidth]{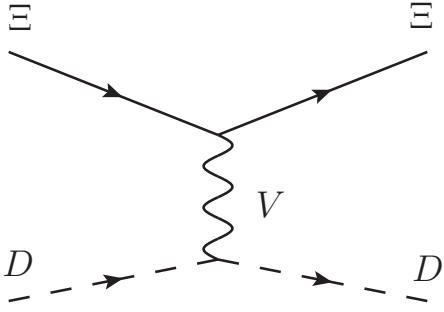}
\caption{Feynman diagram for one-particle exchange $D\Xi$ interaction.}\label{Feynman1}
\end{figure}
In this work, we describe the $D\Xi$ interaction by one-particle exchange diagrams as shown in Fig~\ref{Feynman1}. In the chiral limit and the heavy quark limit, the effective Lagrangian of the interacting vertices involved are \cite{Feng:2011zzb,Liu:2001ce}
\begin{eqnarray}
\mathcal{L}_{DDV}
  &=&-\mathrm i g_{DDV} D_i^{\dag} \stackrel{\leftrightarrow}{ \partial}_\mu D_j
\mathbb{V}^{ij\mu},\nonumber\\
\mathcal{L}_{\Xi\Xi V}&=&- g_{\Xi\Xi V} \bar \Xi_i\gamma_\mu\mathbb{V}^{ij\mu} \Xi_j,
\end{eqnarray}
where $g_{DDV}$ and $g_{\Xi\Xi V}$ are the coupling constants, and $\mathbb{V}$ refers to the fields of the vector mesons and have the following form:
\begin{eqnarray}
\mathbb{V}=\left(\begin{array}{ccc}
\frac{\rho^{0}}{\sqrt{2}}+\frac{\omega}{\sqrt{2}}&\rho^{+}\\
\rho^{-}&-\frac{\rho^{0}}{\sqrt{2}}+\frac{\omega}{\sqrt{2}}
\end{array}\right),\nonumber
\end{eqnarray}
where the $\phi$ meson exchange process is neglected because of the OZI suppression. In order to include the finite-size effects of these hadrons, we also introduce a form factor at each interacting vertex of hadrons. Following Refs.~\cite{Guo:2007mm,Xie:2010zza,Feng:2011zzb}, we take the monopole form:
\begin{eqnarray}\label{FF}
F(k^2)=\frac{\Lambda^2-m_V^2}{\Lambda^2-k^2},
\end{eqnarray}
where $m_V$ is the mass of the exchanged meson, $k$ is the momentum transfer carried by the exchanged meson, and $\Lambda$ is a cutoff parameter. In the instantaneous approximation, we have $k=k_t=p_t-q_t$.

The BS equation can be treated in the so-called ladder approximation \cite{DL}. In this approximation, $K(P,\,p,\,q)$ is replaced by its lowest order form. Using these interaction vertices and the form factor, one can get the following kernel:
\begin{eqnarray}
K(P,\,p,\,q) = \sum_{V=\rho,\,\omega} \frac{-\mathrm i g_{DDV}\cdot g_{\Xi\Xi V}\cdot c_V \left[F(k_t^2)\right]^2}{(q_t-p_t)^2-m_V^2}\hspace{3mm} \nonumber\\[5pt]
\cdot \Big[2(\lambda_2M+p_l)v\!\!\!/+p\!\!\!/_t+q\!\!\!/_t-\frac{1}{m_V^2} (q_t^{2}-p_t^2)(q\!\!\!/_t-p\!\!\!/_t)
 \Big],\label{Kernel}
\end{eqnarray}
where we have used the covariant instantaneous approximation \cite{Guo:1996jj,Guo:2007mm,Xie:2010zza,Feng:2011zzb,Wang:2017rjs}, $p_l=q_l$, and $c_{\rho(\omega)}=\frac{1}{2}(-\frac{1}{2})$ and $\frac{3}{2}(\frac{1}{2})$ corresponding to isoscalar and isovector mesons, respectively.

In the chiral limit and the heavy quark limit, the propagator of the $D$ meson can be expressed at the leading order of the $1/m_D$ expansion as follows \cite{Feng:2012zzf}:
\begin{eqnarray}\label{SD}
s_D(\lambda_2P-p)&=&\frac{\mathrm i}{2m_2(p_l+M-m_2+\mathrm i \epsilon)}.
\end{eqnarray}
The propagator of $\Xi$ has the form \cite{Liu:2015qfa,Wang:2017rjs}
\begin{eqnarray}\label{SXi}
s_\Xi(\lambda_1P+p)&=&\frac{\mathrm i \left [(\lambda_1M+p_l) v\!\!\!/+p\!\!\!/_t+m_1\right]}{(\lambda_1M+p_l-\omega_1+\mathrm i \epsilon)(\lambda_1M+p_l+\omega_1-\mathrm i \epsilon)}.\nonumber\\[2pt]
\end{eqnarray}

Then, we substitute Eqs.~\eqref{BSA},\,\eqref{Kernel}--\eqref{SXi} into Eq.~\eqref{BSES} and obtain the BS equation for the $D\Xi$ bound state:
\begin{eqnarray}\label{BSEF4}
&&f(p_l,\,p_t)\nonumber\\[2pt]
&=&\sum_{V=\rho,\,\omega} \frac{\mathrm i g_{DDV}\cdot g_{V\Xi\Xi}\cdot c_V\left [(\lambda_1M+p_l) v\!\!\!/+p\!\!\!/_t+m_1\right]}{2m_2(p_l+M-m_2+\mathrm i \epsilon)} \nonumber\\[6pt]
&&\quad \times \frac{1}{(\lambda_1M+p_l-\omega_1+\mathrm i \epsilon)(\lambda_1M+p_l+\omega_1-\mathrm i \epsilon)}\nonumber\\[6pt]
&&\quad \times \int \frac{{\mathrm d}^4 q}{(2\pi)^4} \left[F(k_t^2)\right]^2 \frac{1}{(q_t-p_t)^2-m_V^2}\Big[
2(\lambda_2M+p_l)v\!\!\!/  \nonumber\\[6pt]
&& \hspace{1.2cm} +p\!\!\!/_t+q\!\!\!/_t-\frac{1}{m_V^2} (q_t^2-p_t^2) (q\!\!\!/_t-p\!\!\!/_t) \Big] f(q_l,\,q_t).\nonumber\\
\end{eqnarray}

In the bound state rest frame, one has $p_t=(0,\,-{\bf p}_t)$, ${\bf p}_t^2=-p_t^2$ and ${\bf p}_t\cdot {\bf q}_t=-p_t\cdot q_t$. Performing the integration over $p_l$ on both sides with the residue theorem, we have
\begin{eqnarray}\label{BSEf}
\tilde{f}({\bf p}_t)&=& -\sum_{V=\rho,\,\omega}\frac{\pi g_{DDV}\cdot g_{\Xi\Xi V}\cdot c_V}{2m_2\omega_1(M-\lambda_1M-\omega_1-m_2)}\nonumber\\[4pt]
&&\quad\times \int \frac{{\mathrm d}^3 {\bf q}_t}{(2\pi)^3} \left[F({\bf k}_t^2)\right]^2 \frac{1}{({\bf q}_t-{\bf p}_t)^2+m_V^2}\nonumber\\[4pt]
&& \qquad\times \bigg\{-2\omega_1\Big[(\lambda_2-\lambda_1)M-\omega_1\Big] -{\bf p}^2_t -{\bf p}_t\cdot {\bf q}_t\nonumber\\
&&\qquad\qquad+\frac{1}{m_V^2} ({\bf q}_t^2-{\bf p}_t^2) ( {\bf p}_t \cdot {\bf q}_t -{\bf p}_t^2   ) \bigg\}\tilde{f}({\bf q}_t),\nonumber\\
\end{eqnarray}
where we have defined the function $\tilde{f}({\bf p}_t)=\int \frac{\mathrm d p_l}{2\pi}f(p_l,\,{\bf p}_t)$. One may note that this equation involves the integration of $q_t$ and it looks like a divergent integration since $q_t$ varies from 0 to $+\infty$. However, the Lorentz-scalar function, $\tilde{f}({\bf p}_t)$, decreases to zero rapidly at the large momentum transfer and thus there is no divergence in practice \cite{Guo:2007mm,Xie:2010zza,Feng:2011zzb,Feng:2011zzb,Wang:2017rjs}. Since we study the ground state of the $D\Xi$ bound state, the BS wave function is in fact rotationally invariant and depends only on the norm of the three momentum, $|{\bf p}_t|$. Then, the BS equation becomes a one-dimensional integral equation.

\subsection{Normalization condition}
In Eq.~\eqref{BSEf}, we leave the normalization undetermined. Following Ref.~\cite{DL}, the normalization condition for the BS equation can be written as
\begin{eqnarray}\label{nc1.2}
-\frac{\mathrm i}{(2 \pi)^4}\int\mathrm d^4 p \mathrm d^4 q \bar \chi_P(p)\frac{\partial }{\partial P_0}\left [I(P,\,p,\,q)+K(P,\,p,\,q)\right ]\nonumber\\[4pt]
\times \chi_P(q)=1,\quad(P_0=E_P),\qquad
\end{eqnarray}
where $I(P,\,p,\,q)$ is the inverse of the four-point propagator
\begin{eqnarray}\label{nc1.2}
I(P,\,p,\,q)=\delta^{(4)}(p-q)[s_\Xi(\lambda_1P + p)]^{-1}[s_D(\lambda_2P - p)]^{-1}.\nonumber\\
\end{eqnarray}

In the $D\Xi$ bound state rest frame, the normalization condition can be written in the following form:
\begin{eqnarray}
-\frac{\mathrm i}{(2 \pi)^4}\int\mathrm d^4 p \mathrm d^4 q \bar \chi_P(p)\frac{\partial }{\partial P_0}[I(P,\,p,\,q)+K(P,\,p,\,q)]\chi_P(q)\nonumber\\[7pt]
=-\frac{8\mathrm i \lambda_1Mm_2}{(2 \pi)^4}\int\mathrm d^4 p (p_l+M-m_2+\mathrm i\epsilon)f^2(p_l,\,{\bf p}_t)=1.\quad\nonumber\\
\end{eqnarray}
According to Eqs.~\eqref{BSEF4} and \eqref{BSEf}, we have
\begin{eqnarray}
f(f_l,\,{\bf p}_t)&=&\frac{\mathrm i}{\pi}\cdot \frac{2\omega_1(\lambda_2M-\omega_1-m_2)}{(p_l+M-m_2+\mathrm i \epsilon)(\lambda_1M+p_l-\omega_1+\mathrm i \epsilon)}\nonumber\\[3pt]
&&\qquad\quad\times\frac{1}{(\lambda_1M+p_l+\omega_1-\mathrm i \epsilon)}\tilde{f}({\bf p}_t).
\end{eqnarray}
Then, One can recast the normalization condition for the BS wave function into the form
\begin{eqnarray}
-\frac{\lambda_1Mm_2}{\pi^5}\int\mathrm d^3 {\bf p}_t \frac{4\omega_1^2(\lambda_2M-\omega_1-m_2)^2}{(\lambda_2M+m_2-\omega_1)^2(\lambda_2M+m_2+\omega_1)^2}\nonumber\\[4pt]
\qquad\qquad \times \tilde{f}^2({\bf p}_t)=1.\qquad
\end{eqnarray}
We note that the unit of $\tilde{f}({\bf p}_t)$ is GeV$^{-5/2}$.

\section{the decay width of the $D \Xi$ bound state}
\begin{figure}[bt]
\centering
\scalebox{0.9}[0.9]{\includegraphics{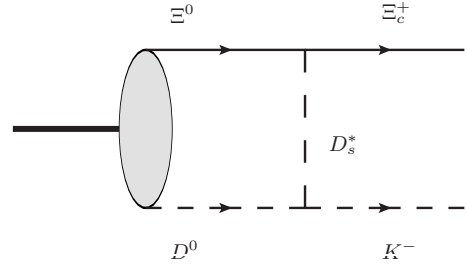}}
\caption{Feynman diagrams for the $D\Xi$ bound states decay into the $\Xi_c^+K^-$ final state.}\label{f2decay}
\end{figure}

In this section, we will proceed to calculate the decay of the $D\Xi$ bound state through BS technique. In the experiments, the LHCb collaboration observed the new $\Omega_c^0$ resonances in the $\Xi^+_c K^-$ spectroscopy \cite{Aaij:2017nav}. In order to compare with the experimental results, we will study the process of the $D \Xi$ bound state decay into above final state. The interaction is described via the vector meson exchange (the $D^*_s$ meson) as shown in Fig.~\ref{f2decay}. The $D^0K^-D_s^*$ coupling vertex is given below in the heavy quark limit and the chiral limit \cite{Yan:1992gz,Wise:1992hn,Burdman:1992gh,Kang:2013jaa}
\begin{eqnarray}
\mathcal{L}_{D^0 K^- D_s^*}
  &=& g_{D^0 K^- D_s^*} (D^{0\dag}D_s^{*\mu}+D_s^{*\mu\dag}D^0)   \partial_\mu K^-.
\end{eqnarray}
The effective Lagrangian for the $\Xi^0 \Xi_c^+ D^*_s $ vertex is
\begin{eqnarray}
\mathcal{L}_{\Xi^0 \Xi_c^+ D^*_s }&=&g_{\Xi^0  \Xi_c^+ D^*_s} \bar \Xi_c^+ \gamma_\mu D^{*\mu}_s \Xi^0+h.c.,
\end{eqnarray}
where SU(3) flavor symmetry is extended to the more general SU(4) symmetry including the charm quark \cite{Hofmann:2005sw}. Considering the charm quark, $\Xi^0$ and $\Xi_c^+$ are collected into SU(3) multiplet fields $8_F$ and $6_F$, respectively, and the $D^*_s$ meson belongs to SU(3) multiplet fields $\bar 3_F$. The coupling strength among baryon $8_F$-plet, $6_F$-plet and meson $\bar 3_F$-plet is
\begin{eqnarray}\label{ggg}
g^{\bar 3}_{86}=\frac{\sqrt 2}{2}g,
\end{eqnarray}
where $g$ is a universal coupling constant in the SU(3) symmetry without the charm quark. One may doubt the validity of this relation with the limit of a light charm quark mass. In Ref.~\cite{Hofmann:2005sw}. the authors demonstrated that Kawarabayashi-Suzuki-Riazuddin-Fayyazuddin (KSRF) relation can be derived by insisting on SU(4) symmetric coupling constants. They found empirically that the relation \eqref{ggg} can provide magnitudes for the coupling reliable within a factor two, i.e. $g^{\bar 3}_{86}=\sqrt 2 g$.

We define $p_a[=(E_a,\,-{\bf p}_a)]$ and $p_b[=(E_b,\,-{\bf p}_b)]$ to be the momenta of $\Xi^+_c$ and $K^-$, respectively. $p^\prime(=\lambda^\prime_2 p_a-\lambda^\prime_1 p_b)$ is defined as the relative momentum between $\Xi^+_c$ and $K^-$ where $\lambda^\prime_{1}=\frac{m_{a}}{m_{a}+m_{b}}$, $\lambda^\prime_{2}=\frac{m_{b}}{m_{a}+m_{b}}$, with $m_{a}$ and $m_{b}$ being the masses of $\Xi^+_c$ and $K^-$, respectively. According to the kinematics of the two-body decay, in the rest frame of the bound state one has
\begin{eqnarray}
E_b&=&\frac{M^2-m_a^2+m_b^2}{2M}, \qquad E_a=\frac{M^2-m_b^2+m_a^2}{2M},\nonumber\\[5pt]
|\vec p_a|&=&|\vec p_b|=\frac{\sqrt{\left[M^2-(m_a+m_b)^2\right]\left[M^2-(m_a-m_b)^2\right]}}{2M}.\nonumber\\
\end{eqnarray}
The differential decay width reads
\begin{eqnarray}
\mathrm d \Gamma=\frac{1}{32 \pi^2}|\mathcal M|^2\frac{|{\bf p}_a|}{M^2}\mathrm d \Omega,
\end{eqnarray}
where $\Omega$ is the solid angle of $\Xi^+_c$, and the amplitude, $\mathcal M$, can be obtained based on the BS techniques \cite{DL,Wang:2017rjs}. We will consider the instantaneous approximation, $p_l= p_l^\prime$, again.

According to the BS technique, the decay amplitude of the isoscalar $D\Xi$ bound state to $\Xi^+_c K^-$ can be written as \cite{Wang:2017rjs}
\begin{eqnarray}
\mathcal M_{I=0}&=&g_{D^0  K^- D^*_s} \cdot g_{\Xi_0 \Xi_c^+ D^*_s }\cdot C_{(0,\,0)}^{21} \nonumber\\[4pt]
 &&\quad \times \int\frac{\mathrm{d}^3 p_t}{(2\pi)^3}F^2(p_t) \bar{u}_{\Xi_c} (p_a)\gamma^\mu f_{I=0}(p_t)u_{\Omega_c}(P) \nonumber\\[4pt]
 &&\qquad \times \left [g_{\mu\nu}+\frac{(p_t^\prime-p_t)_{\mu}(p_t^\prime-p_t)_{\nu}}{m^2_{D^*_s}}\right ](p_2+p_b)^\nu,\nonumber\\
\end{eqnarray}
where $p_t^\prime[=p^\prime-(v\cdot p^\prime)v]$ is the relative transverse momentum between $\Xi^+_c$ and $K^-$, $P(=mv)$ is the momentum of the $D\Xi$ bound state, $p_t$ is the relative transverse momentum between $D$ and $\Xi$ in the bound state, $u_{\Omega_c}$ and $u_{\Xi_c^+}$ and are the Dirac spinors for the bound state $\Omega_c(3188)$ and $\Xi_c^+$, respectively, and $m_{D_s^*}$ is the mass of $D^*_s$.

For the isovector channel, considering the $\Xi^+_c K^-$ final state, we calculate the $I_3=0$ state decay and the corresponding amplitude is
\begin{eqnarray}
\mathcal M_{I=1}&=&g_{D^0  K^- D^*_s} \cdot g_{\Xi_0 \Xi_c^+ D^*_s }\cdot C_{(1,\,0)}^{21} \nonumber\\[4pt]
 &&\quad \times \int\frac{\mathrm{d}^3 p_t}{(2\pi)^3}F^2(p_t) \bar{u}_{\Xi_c} (p_a)\gamma^\mu f_{I=1}(p_t)u_{\Omega_c}(P)\nonumber\\[4pt]
  &&\qquad \times \left [g_{\mu\nu}+\frac{(p_t^\prime-p_t)_{\mu}(p_t^\prime-p_t)_{\nu}}{m^2_{D^*_s}}\right ](p_2+p_b)^\nu.\nonumber\\
\end{eqnarray}

\section{Numerical result}
\begin{table*}[tb]
\renewcommand\arraystretch{1.2}
\centering
\caption{\vadjust{\vspace{-5pt}}The values of $\Lambda$ and corresponding $M$ for which the non-trivial solutions exist for the isoscalar $D\Xi$ bound state.}\label{LMI}
\begin{tabular*}{\textwidth}{@{\extracolsep{\fill}}cccccc}
\hline
\hline
 $M$\,(MeV)    &3100&3125&3150&3175&3200 \\
 \hline
$\Lambda$\,(GeV) &1.423&1.445&1.452&1.439&1.437 \\
\hline
\hline
\end{tabular*}
\end{table*}

\begin{table*}[tb]
\renewcommand\arraystretch{1.2}
\centering
\caption{\vadjust{\vspace{-5pt}}The values of $\Lambda$ and corresponding $M$ for which the non-trivial solution exist for the isovector $D\Xi$ bound state.}\label{LMII}
\begin{tabular*}{\textwidth}{@{\extracolsep{\fill}}cccccc}
\hline
\hline
 $M$\,(MeV)    &3100&3125&3150&3175&3200 \\
\hline
$\Lambda$\,(GeV) &2.910&2.917&2.928&2.933&2.933  \\
\hline
\hline
\end{tabular*}
\end{table*}

In this part, we will solve the BS equation numerically and study whether the $S$-wave $D\Xi$ bound state exists or not.

Since the BS wave function for the ground state is in fact rotationally invariant, $\tilde{f}$ depends only on $|p_t|$. Generally, $|p_t|$ varies from 0 to $+\infty$ and $\tilde{f}$ would decrease to zero when $| p_t|\to +\infty$. We replace $|p_t|$ by the variable, $t$:
\begin{eqnarray}
|p_t|=\left[\epsilon+50 \cdot \ln\left(1+\frac{1+t}{1-t}\right)\right]\,\mathrm {MeV},
\end{eqnarray}
where $\epsilon$ is a small parameter and is introduced to avoid divergence in numerical calculations and $t$ varies from -1 to 1. We then discretize Eq.~\eqref{BSEf} into $n$ pieces ($n$ is large enough) through the Gauss quadrature rule. The BS wave function can be written as $n$-dimension vectors, $f^{(n)}$. The coupled integral equation becomes a matrix equation $f^{(n)}=A^{(n\times n)}\cdot f^{(n)}$ [$A$ corresponds to the coefficients in Eq.~\eqref{BSEf}].

Suppose now that the bound state may exist and its mass is $M$. One gets the numerical results by solving the eigenvalue equation obtained from the matrix equation. When the solution of the matrix equation exists, i.e. the eigenvalue is 1.0, we say that $D\Xi$ forms a $S$-wave ground bound state with the mass being $M$.

In principle, the cutoff parameter $\Lambda$ in our model is not a free parameter. It contains the information of the non-point interaction among the hadrons. However, $\Lambda$ cannot be exactly determined and depends on the specific process in practice. It should be the typical hadronic scale, which is about 1\,GeV. In Ref.~\cite{Liu:2001ce} one takes $\Lambda_{\pi NN}=1.3$\,GeV and $\Lambda_{\rho NN}=1.4$\,GeV. For the mesons with heavy quarks, the value of $\Lambda$ can be as large as 3\,GeV \cite{Zhang:2006ix}. $\Lambda$ is varied in the ranges 0.8-4.5\,GeV \cite{Xie:2010zza} and 1-4\,GeV \cite{Feng:2011zzb} when studing the $DK$ bound state. In the present paper, we vary $\Lambda$ from 1 to 5\,GeV.

\begin{figure}[bt]
\centering
\includegraphics[bb=100 260 568 562, clip=true, width=0.5\textwidth]{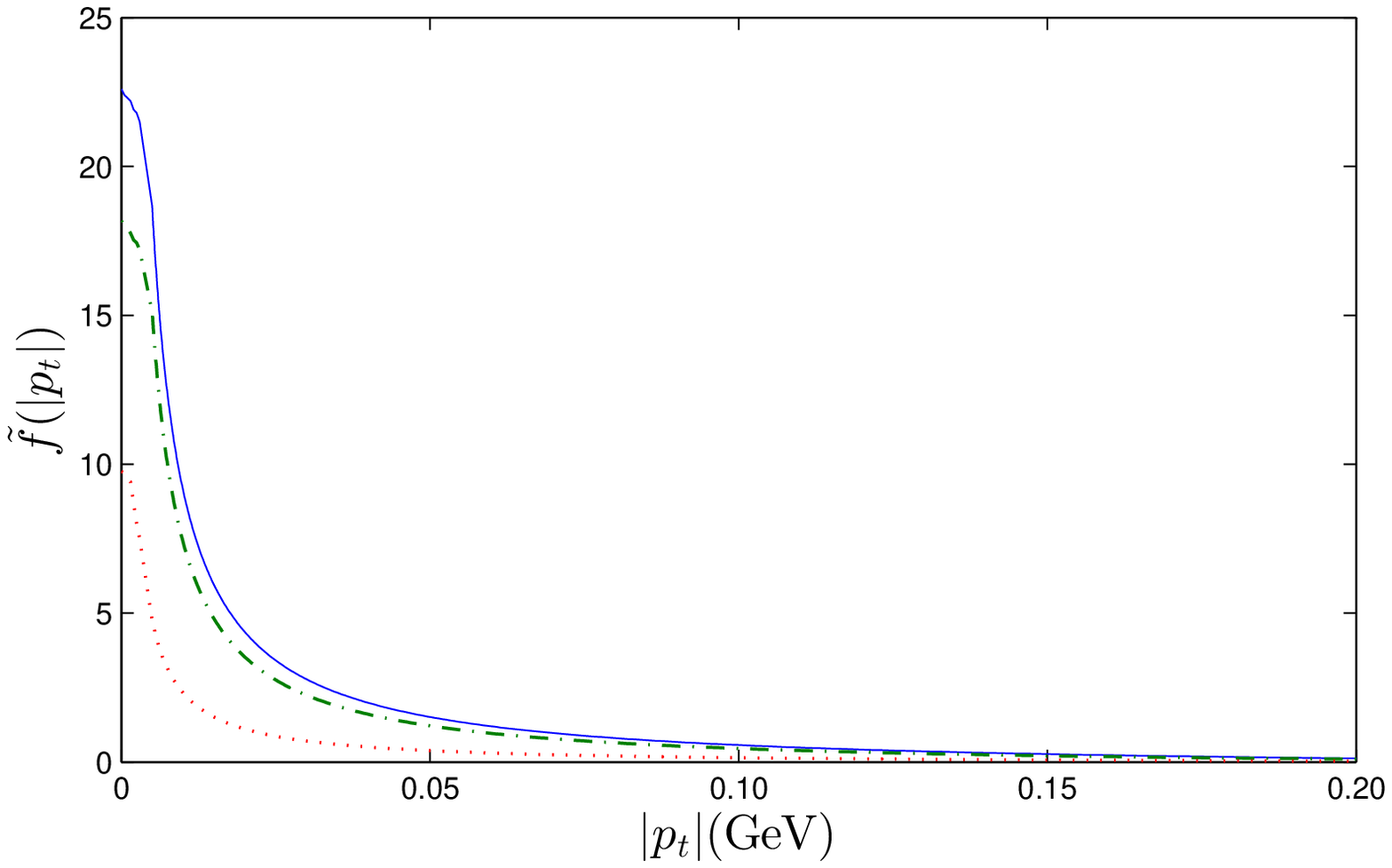}
\caption{Numerical results for the BS wave functions of the isoscalar bound state. The dot line, dot-dash line, and solid line correspond to the cases when $\Lambda$= 1.439, 3.734, and 4.028GeV, respectively.}\label{f1wave}
\end{figure}

In our calculation, we take the mass of the mesons and baryons from the Review of PDG \cite{Patrignani:2016xqp}. There are several coupling constants in our calculations. Following Ref.~\cite{Feng:2012zzf}, we have $g_{DDV}=\beta g_V/\sqrt 2$ where $g_V$ is determined by the KSRF relation, $g_V=m_\rho/f_\pi(\backsimeq5.8)$ and $\beta=\sqrt 2m_V/(g_V f_V)(\backsimeq0.9)$. The coupling constants for the $\Xi\Xi V$ vertex in Fig.~1 and the $D^0K^-D^*_s$ vertex in Fig.~2 are obtained in the framework of light-cone QCD sum rules, $g_{\Xi\Xi V}=1.5$ \cite{Wang:2007yt}, and $g_{D^0K^-D^*_s}=1.84$ \cite{Wang:2006ida}. As for the coupling constant $g_{\Xi^0  \Xi_c^+ D^*_s}$, we estimate its value by extending to the more general SU(4) symmetry including the charm quark as mentioned in Sect.~IV. We take $g_{\Xi^0  \Xi_c^+ D^*_s}=\frac{2}{\sqrt 6}g^{\bar 3}_{86}(\simeq7.6)$ with $g\simeq6.6$ \cite{Hofmann:2005sw}.

Then, we vary the mass of the possible bound state, $M$, from 3100\,MeV to 3200\,MeV and look for the proper values of M and $\Lambda$ for which the non-trivial solutions of the eigenvalue equation exist. Let us first consider the isoscalar case. We find that there are three values of $\Lambda$ corresponding to each $M$. In Table~\ref{LMI}, we list the minimum values of $\Lambda$ and the corresponding situations when $M=3100$, 3125, 3150, 3175, 3200\,MeV. When the values of $\Lambda$ is less than the minimum values, there is no numerical solution of the BS wave function, i.e., $D\Xi$ cannot form a bound state. In order to study the $\Omega_c(3188)$ structure, we keep the mass of the bound state fixed ($M=3188\,$MeV) and find that there are three values of $\Lambda$, 1.439, 3.734, and 4.028\,GeV, for which we can get the solutions. The numerical results of the BS wave functions are displayed in Fig.~\ref{f1wave}. Then, we apply the numerical solutions of the BS wave function corresponding to $M=3188\,$MeV to calculate the width of $D\Xi$ decay into $\Xi_c^+K^-$. The decay widths are given below
\begin{align}
\Gamma^{I=0}&=900\,\text{keV},&\text{when}  ~\Lambda=1.439\,\text{GeV}, &\nonumber\\
\Gamma^{I=0}&=10\,\text{MeV},&\text{when}  ~\Lambda=3.734\,\text{GeV} ,&\nonumber\\
\Gamma^{I=0}&=15\,\text{MeV},                 &\text{when} ~\Lambda=4.028\,\text{GeV}.&\nonumber
\end{align}

For the isovector bound state, we list the minimum values of $\Lambda$ and corresponding $M$ in Table~\ref{LMII}. When keeping the mass of the bound state fixed ($M=3188\,$MeV), we can get two values of $\Lambda$, 2.933 and 4.254\,GeV. The
numerical results of the BS wave functions are displayed in Fig.~\ref{f2wave}. The corresponding decay widths are
\begin{align}
\Gamma^{I=1}&=3\,\text{MeV},&\text{when}  ~\Lambda=2.933\,\text{GeV}, &\nonumber\\
\Gamma^{I=1}&=32\,\text{MeV},&\text{when}  ~\Lambda=4.254\,\text{GeV}.&\nonumber
\end{align}

According to the LHCb collaboration experiments, the peak and width of $\Omega_c(3188)$ is $3188\pm{5}\pm{13}\,$ and $60\pm15\pm11$\,MeV, respectively \cite{Aaij:2017nav}. We can see the two $D\Xi$ bound states in our calculations are located in the range of $\Omega_c(3188)$ and its decay width is smaller than that of $\Omega_c(3188)$. That is to say, the $D\Xi$ bound states are likely to exist and could contribute to the observed $\Omega_c(3188)$. However, $\Omega_c(3188)$ might also be produced by a superposition of others resonances. We still need more experimental data to support our results. If $\Omega_c(3188)$ corresponds to the $S$-wave $D\Xi$ bound states, its quantum numbers should be $I(J^P)=0(\frac{1}{2}^-)$ or $1(\frac{1}{2}^-)$ and one can verify this resonance by its quantum numbers. This is important to ascertain the structure of $\Omega_c(3188)$.

\begin{figure}[bt]
\centering
\includegraphics[bb=15 268 530 589, clip=true, width=0.5\textwidth]{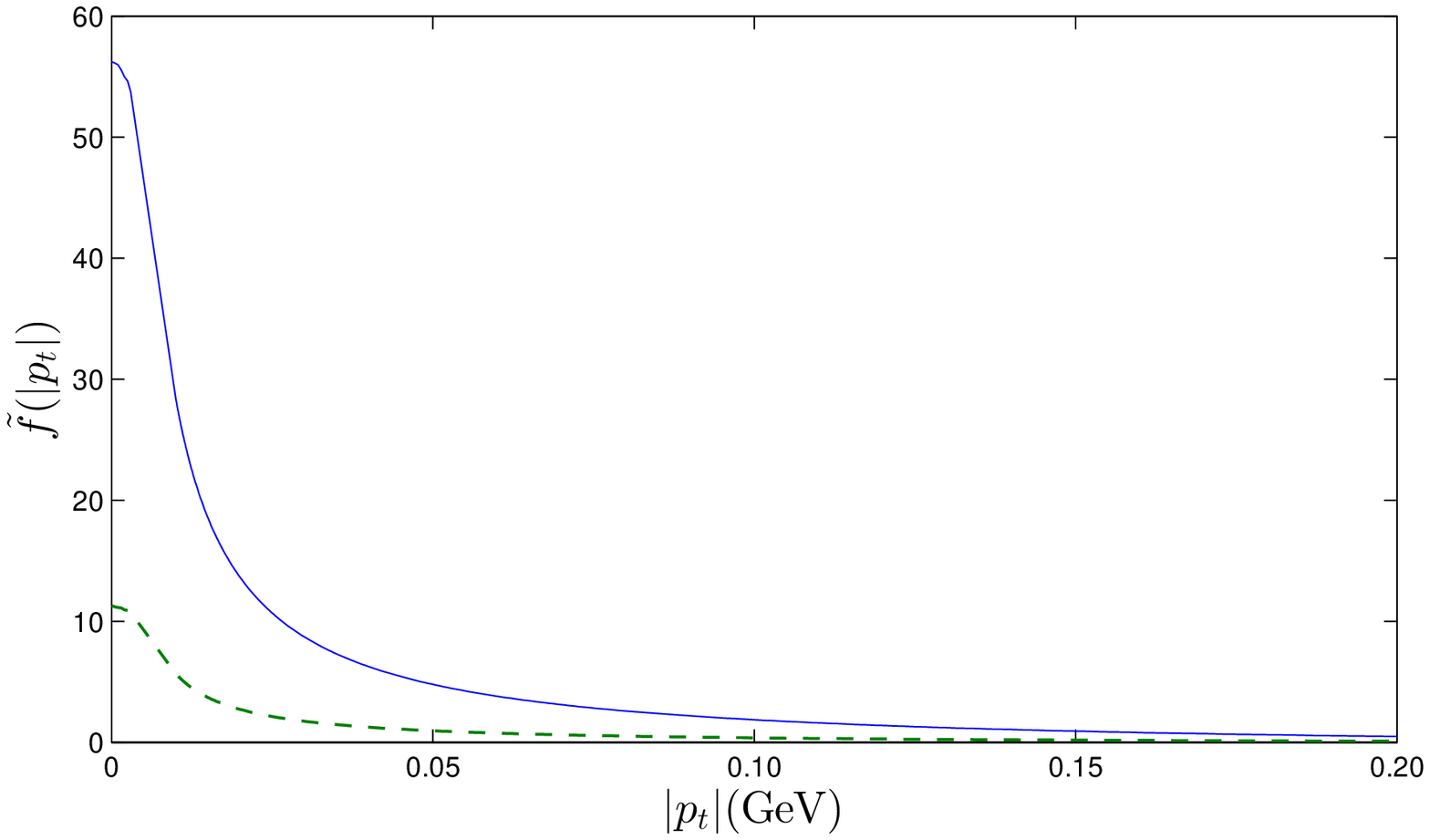}
\caption{Numerical results for the BS wave functions of the isovector bound state. The dash line and solid line correspond to the cases when $\Lambda$=2.933 and 4.254GeV, respectively.}\label{f2wave}
\end{figure}

\section{Summary and discussion}
In this paper, we studied the possible $S$-wave molecular bound states of the $D\Xi$ system in the BS formalism. Considering the interaction kernel based on vector meson exchange diagrams, we established the BS equation for the $D\Xi$ system in the ladder and instantaneous approximations. Then, we discretized the integral equation and solved the eigenvalue equation numerically. We confirmed the existence of the $S$-wave isoscalar and isovector $D\Xi$ bound states in this formalism and obtained their BS wave functions. We also calculated the decay widths of the bound states by using the BS wave functions. According to our results, the $D\Xi$ bound states are compatible with the new observed $\Omega_c(3188)$ structure. Or, at least, we can say safely that these two bound states could contribute to the $\Omega_c(3188)$ peak. Further experimental data are expected to verify $\Omega_c(3188)$ to be the $D\Xi$ bound state with its quantum numbers determined.

We studied the $D\Xi$ bound states since the sum of $D$ and $\Xi$ is just near the 3188\,MeV threshold. One may note that other channels, including $\pi^0 \Omega_c$, $\eta \Omega_c$, $K^-\Xi_c^+$, $K^0\Xi_c^0$, and $D^+_s\Omega^-$et al., would couple to the $D\Xi$ system in the meson-baryon scattering with the strangeness $S=2$ and the charm number $C=1$. The $D_s^+\Omega^-$ state is forbidden to couple to the $S$-wave $D\Xi$ system by its spin quantum number. For other channels including light meson when they couple to $D\Xi$, a charm meson should be exchanged. Then the coupled channel effects will be suppressed in comparison with the $\rho$($\omega$) exchange $D\Xi$ interaction. Therefore, we omitted the coupled channel effects in our calculation.

In this work, we predicted the existence of the $D\Xi$ bound states and discussed the possibility of $\Omega_c(3188)$ to be the $D\Xi$ bound states. Unfortunately, there are several parameters not being well determinated in the calculations. These uncertainties reduced the predictability of our model. In order to describe non-pointlike effects, we introduced a cutoff parameter $\Lambda$. This cutoff parameter is dominated by nonperturbative QCD and cannot be determined at present, which leads to large uncertainties. For a fixed $M$, we vary $\Lambda$ from 1 to 5\,GeV in practice. We find that the solutions exist with three (two) values of $\Lambda$ for a fixed $M$ in the case of isoscalar (isovector) bound state. We cannot determine $\Lambda$ exactly at present and thus give a range of $\Lambda$, which causes the uncertainties of the results. However, there are only three (two) different values of $\Lambda$ for a fixed $M$. Hence our results do not rely on $\Lambda$ heavily. With more experimental data available in the future, we will be able to reduce the uncertainty of $\Lambda$ further.

\section{ACKNOWLEDGEMENT}

Chao Wang is grateful to Zhen-Yang Wang for helpful discussions. This work was supported by the Fundamental Research Funds for the Central Universities of China (Project No.~31020170QD052), NSFC-Yunnan United Fund (Project No.~U1302267), the National Natural Science Foundation of China (Projects No. 11275025, No. 11575023, and No. 11775024), and the National Science Fund for Distinguished Young Scholars (Project No.~31325005).

%
%

\end{document}